\documentclass{optica-article}
\journal{oe}

\usepackage{textcomp}
\usepackage{bm}

\begin{document}

\title{Using an acousto-optic modulator as a fast spatial light modulator}

\author{Xialin Liu,\authormark{1,2} Boris Braverman,\authormark{1,*} Robert W. Boyd,\authormark{1,3,$\dagger$}}

\address{\authormark{1}Department of Physics and Max Planck Centre for Extreme and Quantum Photonics, University of Ottawa, 25 Templeton Street, Ottawa, ON, K1N 6N5, Canada\\
\authormark{2}Key Laboratory of Space Active Opto-electronics Technology, Shanghai Institute of Technical Physics, Chinese Academy of Sciences, 500 Yutian Road, Shanghai 200083, China\\
\authormark{3}Department of Physics and Astronomy, University of Rochester, 275 Hutchison Road, Rochester, NY, 14627, USA.}

\email{\authormark{*}bbraverm@uottawa.ca}
\email{\authormark{$\dagger$}robert.boyd@uottawa.ca}

\begin{abstract*}
High-speed spatial light modulators (SLM) are crucial components for free-space communication and structured illumination imaging. Current approaches for dynamical spatial mode generation, such as liquid crystal SLMs or digital micromirror devices, are limited to a maximum pattern refresh rate of 10 kHz and have a low damage threshold. We demonstrate that arbitrary spatial profiles in a laser pulse can be generated by mapping the temporal radio-frequency (RF) waveform sent to an acousto-optic modulator (AOM) onto the optical field. We find that the fidelity of the SLM performance can be improved through numerical optimization of the RF waveform to overcome the nonlinear effect of AOM. An AOM can thus be used as a 1-dimensional SLM, a technique we call acousto-optic spatial light modulator (AO-SLM), which has 50 \textmu{}m pixel pitch, over 1 MHz update rate, and high damage threshold. We simulate the application of AO-SLM to single-pixel imaging, which can reconstruct a 32$\times$32 pixel complex object at a rate of 11.6 kHz with 98\% fidelity. 
\end{abstract*}

\section{Introduction}

The generation and manipulation of spatial modes of light is desired in tasks such as free-space communication, structured illumination microscopy, and high-dimensional quantum key distribution\cite{forbes2016creation,wang2012terabit,liu2018single,walborn2004entanglement}. Modes of light more complex than ordinary Gaussian beams are most often realized using spatial light modulators (SLMs). While SLMs are versatile and have a multitude of applications, all commercially available SLMs rely on the tunable birefringence of liquid crystal molecules, which results in two major practical drawbacks. First, the refresh rate is low: it takes a time interval on the order of 1 ms to update the generated spatial mode\cite{Turtaev:17,braverman2020fast}. Even using a faster alternative, a digital micromirror device (DMD)\cite{shin2015active,shin2016optical}, the fastest refresh speed is only about 22 kHz\cite{Turtaev:17}, which is limited by the motional speed of the circuit-driven micro-mirror. Second, liquid crystal molecules have a low damage threshold (hundreds of \textmu{}J/$\rm{cm}^2$) and high losses\cite{lazarev2012lcos,Mukohzaka:94,igasaki1999high}. These limitations are especially problematic for pulsed or UV lasers that are commonly used in photolithography and nonlinear optics.

Speeding up the operation of SLMs is of great interest to both fundamental and applied research\cite{moddel1989high,becker2010high}. The emergence of high-speed light modulators has promoted the rapid development of single-pixel imaging (SPI)\cite{duarte2008single}, which reconstructs an image from a set of structured illumination patterns and a single-pixel detector. Although SPI can be applied in contexts where detection technologies are not well-developed\cite{Aspden:15,pelliccia2016experimental,stantchev2020real}, SPI suffers from low imaging rates. For example, reconstructing a 100$\times$100 pixel image requires 10000 measurements for full sampling. If the acquisition for each illumination pattern takes 1 ms, the total image acquisition time is already 10 s. During such a long imaging time, the sample may be damaged by laser irradiation or both its position and shape might change, causing imaging failure. Studies have attempted to reduce acquisition time through compressed sensing\cite{PhysRevA.78.061802,duarte2008single} and intensity estimation from limited photon detection\cite{Liu2018Fast}, but the speed of SPI is still fundamentally limited by the refresh rate of the SLM. Compressed imaging with pattern modulation rates in excess of 1 MHz has been demonstrated\cite{hahamovich2021single} using a spinning mask, resulting in an image acquisition rate of 72 Hz with 101 $\times$ 103 pixels resolution. One drawback of the approach in \cite{hahamovich2021single} is that the illumination patterns are fixed and are always generated in the same order, preventing the use of adaptive or compressive sampling techniques. The development of an SLM with higher efficiency and damage threshold would therefore be particularly valuable for imaging over longer distances and in more complex environments.
 
Acousto-optic modulators (AOMs), which consist of a crystal driven by a piezoelectric transducer, can potentially be a solution for the aforementioned issues. AOMs offer both high speed and high damage threshold; they are most commonly used to control the frequency or propagation direction of light. 

The rapid control offered by AOMs enables the generation of complex illumination patterns in a time-averaged sense, where discrete Gaussian spots are generated in turn. 

In the context of generating and controlling spatially structured light, AOMs have been applied as rapidly tunable cylindrical lenses\cite{kaplan2001acousto,konstantinou2016dynamic}, as a source for synthesized aperture radar\cite{haney1988real}, and to generate a limited set of complex light patterns through the interference of two Gaussian beams\cite{Baum2015}. Previous works regarding fast spatial mode generation and detection based on AOM has observed switching rates for mode generation of up to 500 kHz\cite{braverman2020fast,radwell2014high}, well in excess of the existing SLM or DMD technology. While these works have demonstrated high modulation and switching rates for the generated light patterns, they simply use AOMs to synthesize rapidly movable focal spots, which are then used to address different regions of a conventional SLM, rather than for directly generating arbitrary complex spatial light fields. 
In a recent result, Treptow et al. implemented a holographic acousto-optic light modulation (HALM) system, which can reconstruct any two-dimensional intensity pattern by applying one acousto-optic deflector (AOD) as a holographic modulator to reconstruct image lines, and another AOD as a line deflector\cite{treptow2021artifact}. In \cite{treptow2021artifact}, the AOD is assumed to be illuminated with a light pulse of duration much greater than the inverse acoustical frequency, resulting in an averaging out of any phase coherence between different diffracted components of the output light beam. This results in the generation of a desired intensity distribution which is primarily suitable for imaging applications. In contrast, we consider the opposite limit of a very short laser pulse, where the impact of the relative phases of different frequency components becomes essential. In this regime, our work demonstrates the possibility of fully and deterministically controlling both the phase and amplitude of the optical field.

The present work demonstrates how an AOM can be directly used as a genuine, ultrafast SLM. The spatial grating formed by the sound waves in the AOM reproduces the incident RF waveform and imprints it nearly perfectly onto the spatial mode of the diffracted light. This technique works only with pulse duration much shorter than the highest frequency of the acoustical wave in the AOM, since the acoustical pattern is a traveling wave. Across the 2 mm probe beam width, we can have more than 40 pixels with independently programmable amplitude and phase, which can be used for arbitrary mode generation. Resolution is defined as the length of a single reprogrammable pixel on the output optical field. We note here that the results of \cite{Baum2015} can be interpreted as the realization of a 2-pixel AO-SLM. We numerically simulated the experimental configuration of light propagating through AOM with an index variation produced by an RF signal. One difficulty of utilizing an AOM directly as an SLM is in the nonlinear response of the light amplitude in terms of the incident RF signal, due to light propagation and diffraction within the AOM crystal. To mitigate this nonlinear response, we develop an iterative optimization algorithm that optimizes the input RF signal to produce desired arbitrary modes of light with higher fidelity. The optimization procedure increases the fidelity from 60\% to 85$\%$ for arbitrary phase patterns with 50 \textmu{}m resolution. The refresh rate of our AO-SLM is limited by the transit time of the sound waves across the beam waist, which is only 600 ns for a 2 mm waist size beam at the speed of sound equal to 3.6 km/s, hundreds and thousands of times faster than DMD and liquid crystal spatial light modulator\cite{Turtaev:17}. We also simulated the application of our AO-SLM on structured-illumination SPI, and the high refresh rate greatly accelerates the imaging time by around 3 orders of magnitude relative to SPI by other traditional spatial light modulation methods, requiring only 86 \textmu{}s to reconstruct a 32$\times$32-pixel complex image at 100 \textmu{}m resolution. 

\section{Methods}

\subsection{AO-SLM Simulation}
\begin{figure}[htbp]
\centering
\includegraphics[width=12cm]{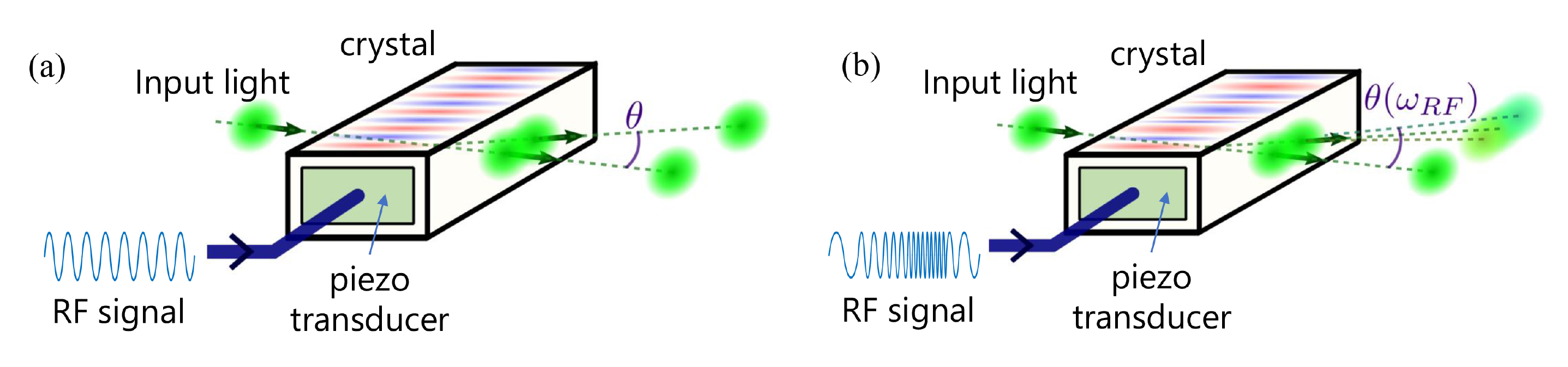}
\caption{(a) Schematic diagram of the AOM with a single-frequency RF input. The majority of the input beam power is diffracted at an angle given by Eq. (\ref{eq:Theta}). (b) Schematic diagram of the AOM with an RF signal containing many frequencies.}
\label{fig:aom}
\end{figure}

The conceptual diagram of an AOM is shown in Fig.\ref{fig:aom}. Fundamentally, an AOM is a crystal driven by a piezoelectric transducer. The sound waves from the electric signal change the index of refraction of the crystal, forming a spatial grating. Typically, the crystal dimensions are around 1 cm, the acoustical velocity is between 500 m/s and 5 km/s, the RF drive frequency is between 50 and 200 MHz, and the RF drive power is around 1 W. Fig.\ref{fig:aom} (a) shows light being diffracted by the grating generated by the sound wave corresponding to an RF waveform of constant amplitude and phase; the different colors inside the crystal represent the compression/stretching of the crystal induced by the piezo transducer. If the amplitude and phase of the input RF signal wave vary with time, there will be a superposition of different RF frequencies inside the AOM and as a result, the light will be diffracted into a superposition of different directions as shown in Fig.\ref{fig:aom} (b). The first-order diffraction angle $\theta$ for a single RF frequency $\omega_{RF}$ can be expressed as,
\begin{equation}
\sin (\theta) =\frac{\lambda}{n\lambda_S} = \frac{\lambda}{n\upsilon_S}\frac{\omega_{RF}}{2\pi}
\label{eq:Theta}
\end{equation}
where $\lambda$ and $\lambda_S$ are the wavelengths of light and sound, $n$ is the refractive index of the crystal, $\upsilon_S$ is the propagation velocity of the sound, and $\omega_{RF}$ is the frequency of RF signal (also the frequency of the sound wave). Ideally, the input beam goes into the AOM at the Bragg angle $\theta_B = \frac{\theta}{2}$, which results in the highest diffraction efficiency. Based on this intuition, one would expect that the diffracted components of the light replicate the spatial refractive index pattern produced by the RF waveform. 

\begin{figure}[htbp]
\centering
\includegraphics[width=10.5cm]{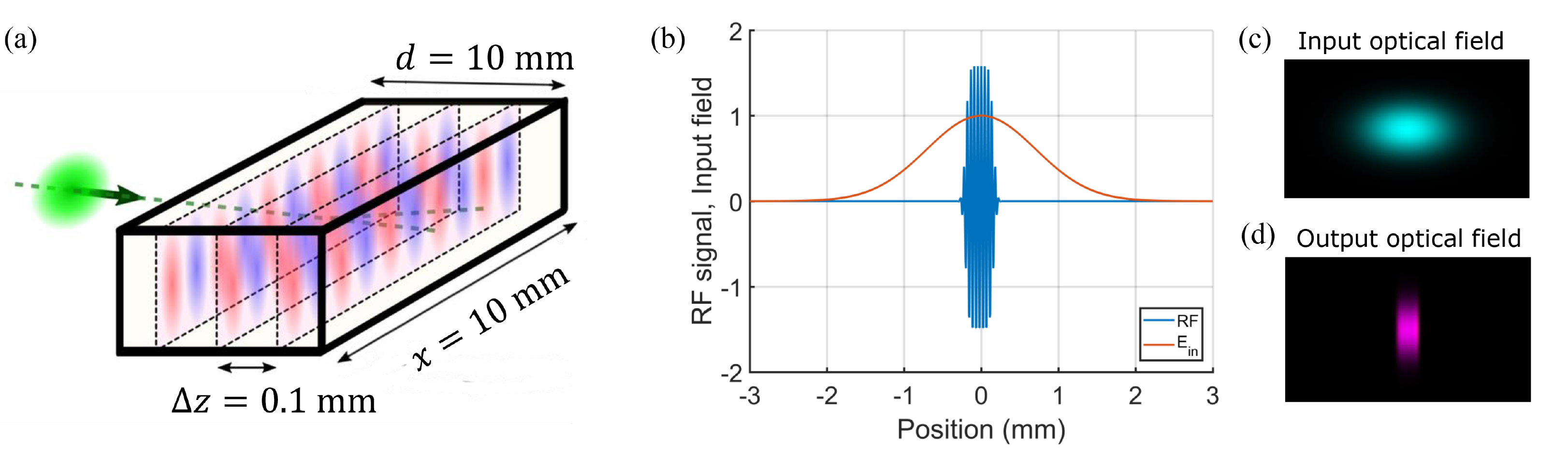}
\caption{(a) Setup for split-step simulation showing typical dimensions of optical crystal in an AOM. (b) Input light field curve (orange), RF signal curve (blue); (c), (d) two-dimensional input and output optical fields.}
\label{fig:aomslm}
\end{figure}

We performed a numerical simulation for the AOM system. Unlike a regular SLM where the light only propagates in the device for around 20 \textmu{}m, the AOM is not "thin" because light experiences significant diffraction as it propagates through the crystal. To accurately model the AOM response, we perform split-step simulation with 5 \textmu{}m spatial resolution and $\Delta z= 0.1$ mm resolution in the z-direction as shown in Fig. \ref{fig:aomslm}(a). That is, we divide the full 10 mm length of the AOM into slices of thickness $\Delta z$. For each slice, we perform a free-space propagation step, and a phase shift induced by the spatial index modulation imprinted by the acoustical wave. The peak index variation induced by the acoustical wave is about $10^{-5}$. The refractive index of the AOM crystal TeO$_2$ is 2.3 at 520 nm wavelength, the speed of sound is $\upsilon_S$ = 3.6 km/s and the central frequency of the RF is set to $\omega_{RF}/{2\pi}$ = 80 MHz, in agreement with the experimental apparatus described later. The fidelity of the output light field relative to the target field (the field imprinted on the RF wave) and the diffraction efficiency are used to evaluate the system performance. Fig. \ref{fig:aomslm}(b) shows a single-slit modulation without optimization. The input optical field is shown by the orange curve and the RF field or the sound wave is shown by the blue curve. We expect that the central part of the optical field to be diffracted, while the remainder of the input optical field is not diffracted. As expected, for an input optical field with the elliptical profile shown in Fig \ref{fig:aomslm}(c), the output field looks as if it has passed through a slit.
\begin{figure}[htbp]
\centering
\includegraphics[width=11.5cm]{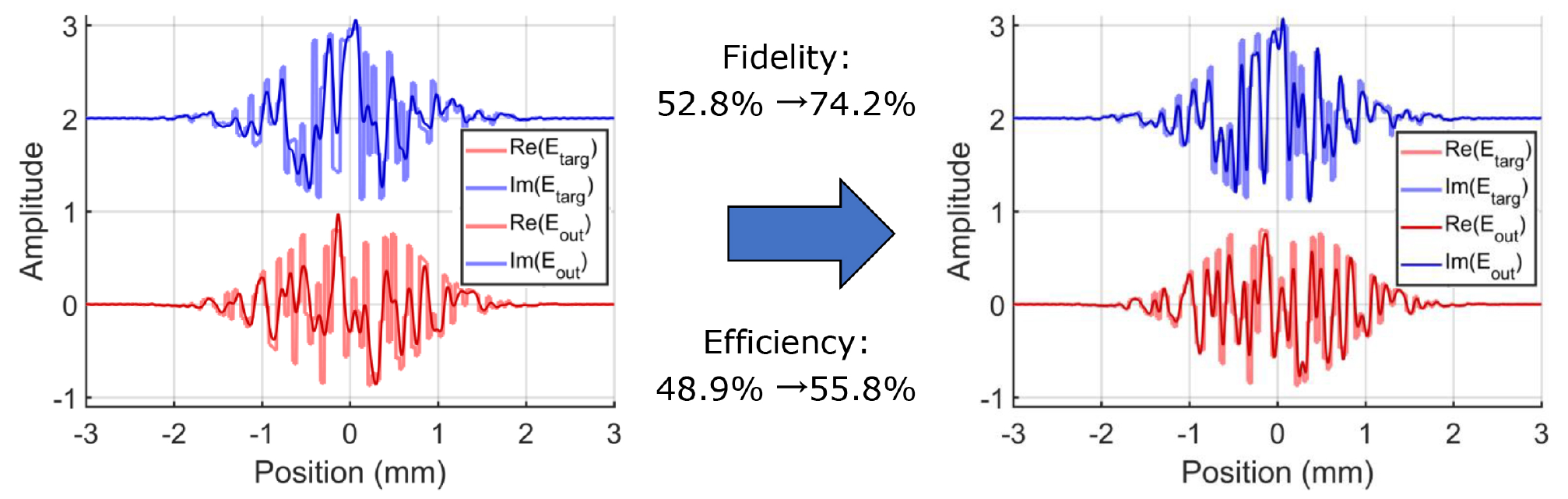}
\caption{Comparison of the optical field without (left) and with (right) optimization. The simulation is performed at 50 \textmu{}m spatial resolution. The imaginary part (blue) is offset vertically by 2 units.}
\label{fig:optimization}
\end{figure}

We can get quite accurate reproductions of the RF waveforms when the RF amplitude and phase do not vary quickly over the laser beam aperture, as shown in Fig. 2 (b). In contrast, for a random pattern with sharp changes of phase like that of Fig. \ref{fig:optimization}, there are many large deviations between the output light field and the target light field and the fidelity without optimization is only 53\%. 

We divide up the AOM (extending over 0 < z < d) into a series of N slabs with thickness $\Delta z$, where $ N = d/ \Delta z$. Then, we perform a split step simulation, alternating between imposing a position-dependent phase (due to the index variation produced by the RF signal $n_{rf}(x)$), and a momentum-dependent phase (corresponding to propagation through the material by distance $\Delta z$) \cite{Turunen1990}. Thus, the overall propagation can be expressed as
 \begin{eqnarray}
E_{out} = [e^{i\Delta z n_{rf}(x)} e^{-i\sqrt{k^2-k_{x}^2}\Delta z}]^N E_{in},
\label{eq:Eout}
 \end{eqnarray}

Note that the two terms are functions of $x$ and $k_x$ respectively, and they do not commute. The term $ n_{rf}(x)$ appears multiple times, being multiplied by itself. Thus, the response of the system will be nonlinear in terms of the function $ n_{rf}(x)$. We optimized the RF pattern, using the Levenberg-Marquardt algorithm by MATLAB to iteratively determine the optimal RF waveform for matching a given target optical field. The optimization procedure can be expressed as

\begin{equation}
\begin{split}
\hat{n}_{rf}(x) = \mathop{\arg\!\min_{n_{rf}(x)}}\frac{1}{2} \left\|E_{target}-E_{out}(n_{rf}(x))\right\|^2_2,\\
= \mathop{\arg\!\min_{n_{rf}(x)}}\frac{1}{2} \sum_{i}\left\|E_{target}(i)-E_{out}(n_{rf}(x),i)\right\|^2_2
\label{eq:LM}
\end{split}
\end{equation}
where $i$ represents $i^{th}$ programmable pixel. 
As seen in Fig 3(b), after this optimization procedure, the optical output is much more similar to the target field, and the fidelity in this example is improved from 53\% to 74\%, even realizing some modest gains in diffraction efficiency. More relative results will be shown in the next section.

\section{Results}
\subsection{Spatial light modulation by AOM}

\begin{figure}[htbp]
\centerline{
\includegraphics[width=13cm]{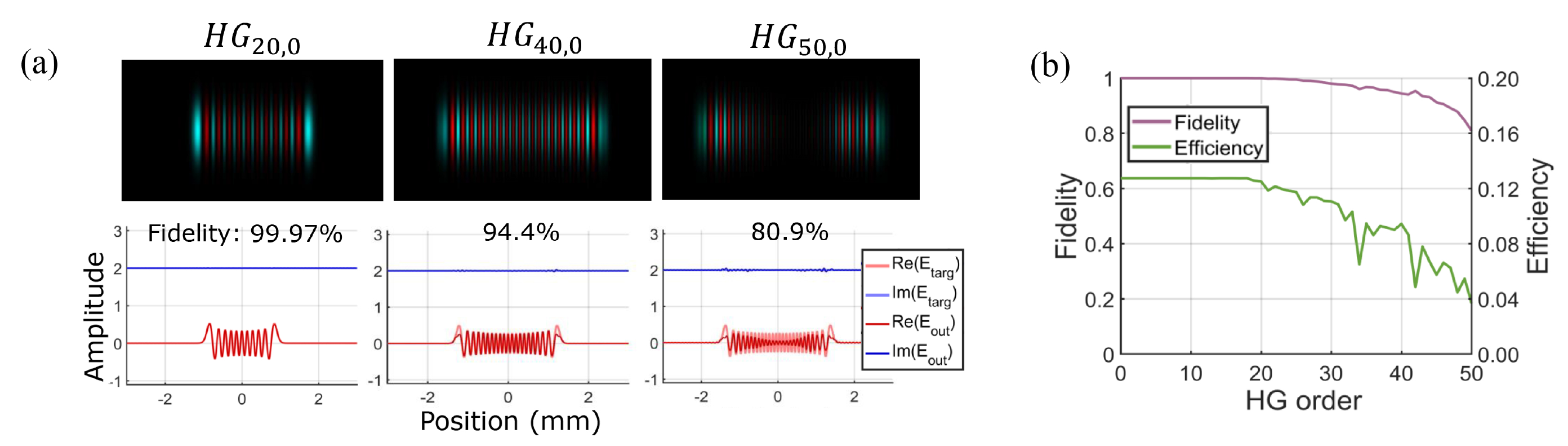}}
\caption{(a) Complex output field of HG modes. The curves at the bottom in dark blue, dark red and light blue, light red respectively represents the real and imaginary parts of the output light field and the target field. The imaginary part (blue) is offset vertically by 2 units. (b) The fidelity and efficiency curves for different orders of HG modes. }
\label{fig:HGmodesSim}
\end{figure}

To verify the feasibility of AO-SLM, we modeled the generation of several different types of spatial light mode. In the simulation we use the AOM as a phase-only device, i.e. to have $E_{out} = E_{in} \times I(x)(e^{i\phi(x)})$, where $A(x)$ and $\phi(x)$ represent the intensity and phase pattern. Fig. \ref{fig:HGmodesSim} shows the simulation results and fidelity and efficiency curves for approximating Hermite-Gaussian (HG$_{n,0}$) modes, which are obtained after RF waveform optimization. The generation of HG modes up to the 30th order maintains high accuracy and efficiency. The efficiency begins to drop significantly after 30 orders, while the fidelity still can reach 80\% at HG$_{50,0}$. We note that to generate HG modes, the diffraction efficiency is necessarily relatively low (at most 12\%) because here we are using a phase-only modulator (such as the AO-SLM) to perform simultaneous amplitude and phase control over a light field. When using the AO-SLM only as a phase modulator (as shown in Fig. \ref{fig:response}(b) below), the diffraction efficiency is correspondingly higher.

\begin{figure}[htbp]
\centerline{
\includegraphics[width=10cm]{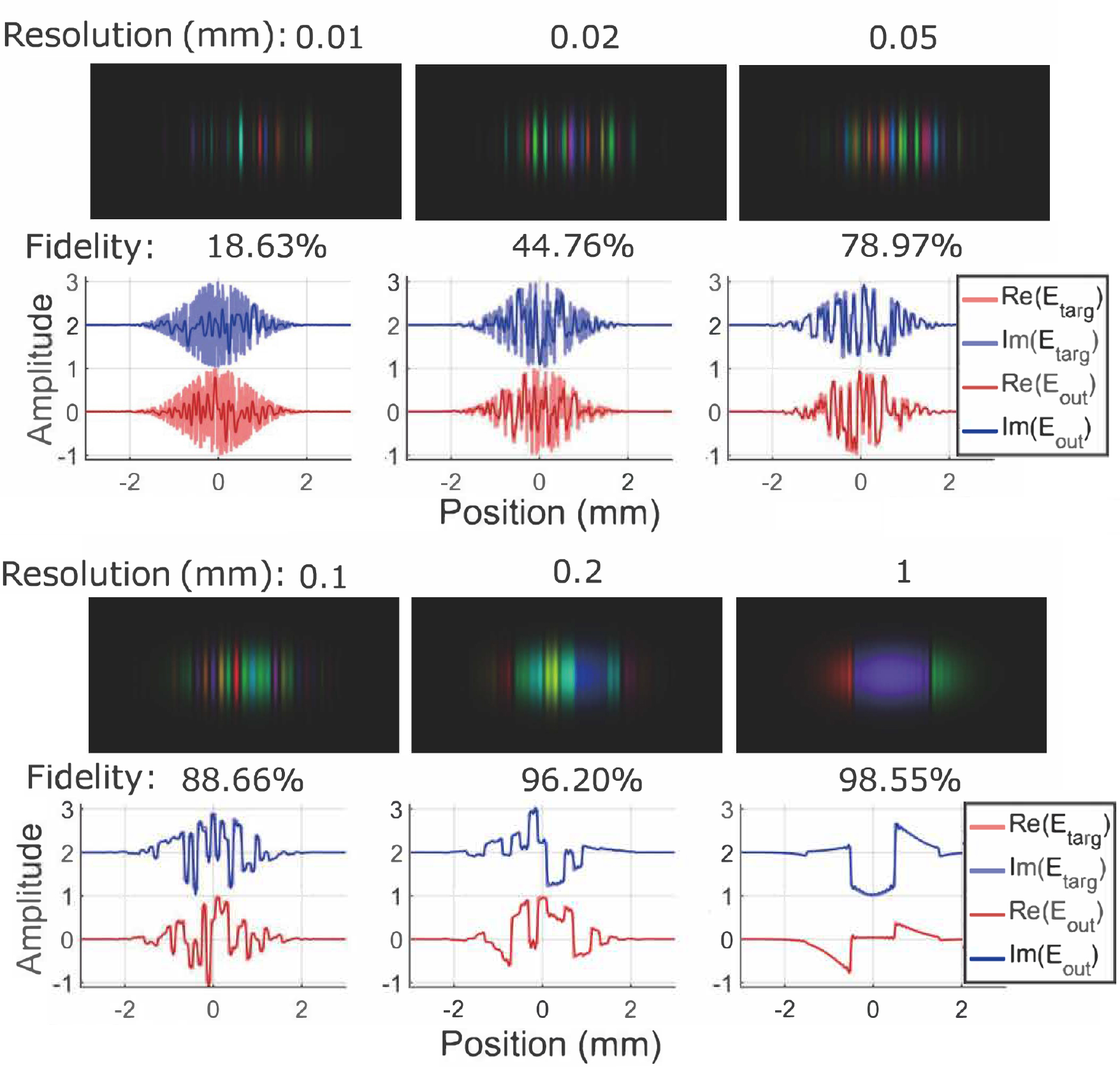}}
\caption{Complex output field for random phase modulation with different spatial resolution ranging from 0.01 to 1 mm. }
\label{fig:RandArraySim}
\end{figure}

\begin{figure}[htbp]
\centerline{
\includegraphics[width=11cm]{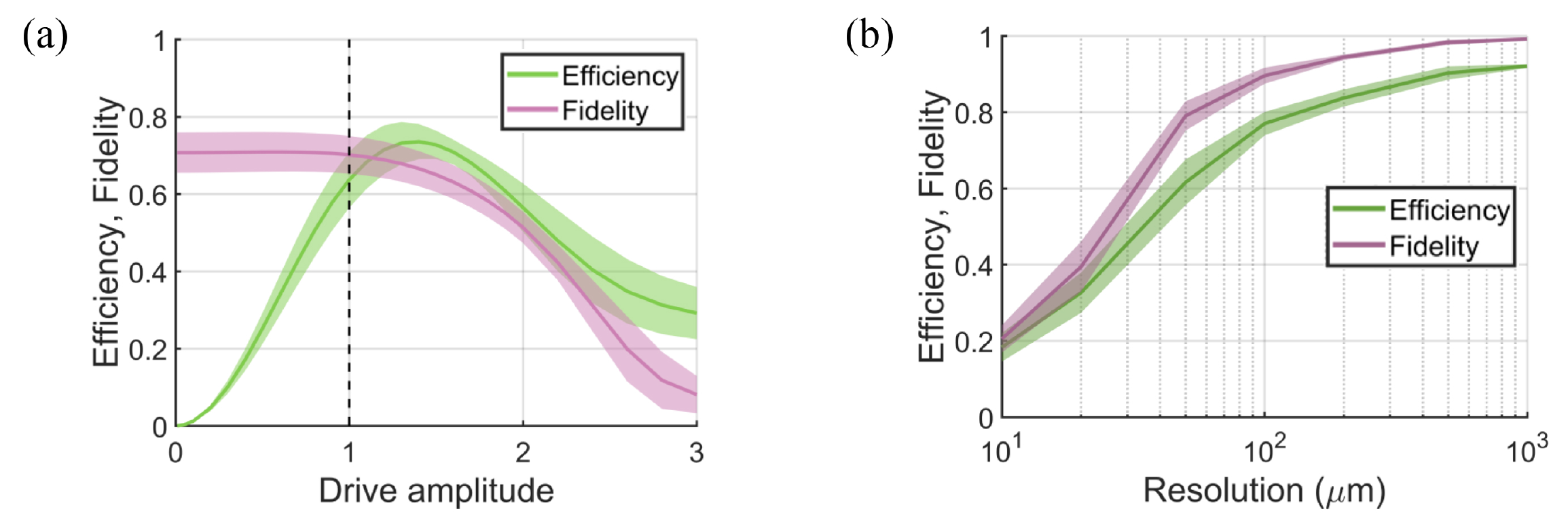}}
\caption{(a) Diffraction efficiency and fidelity for different RF amplitudes with random phase modulation at 45 \textmu{}m resolution. (b) Diffraction efficiency and fidelity curve for different resolutions. These plots correspond to 100 simulations at each point, with the lines indicating the mean value, while the shaded regions correspond to $\pm 1$ standard deviation.}
\label{fig:response}
\end{figure}

 Fig. \ref{fig:RandArraySim} shows the simulation results for random phase modulation with different spatial resolutions. In this simulation, we use an AOM as a phase-only spatial light modulation device, i.e. to have $E_{out} = E_{in} \times (e^{i\phi(x)})$, where the phase pattern, $\phi(x)$, was randomly sampled over the (0,2$\pi$) range. The mode generation fidelity is larger for lower spatial resolutions with larger effective pixel sizes. We can obtain a fidelity of 75\% for spatial 50 \textmu{}m resolution.

Fig. \ref{fig:response} (a) shows the fidelity and efficiency of this system under different RF drive amplitudes at 45 \textmu{}m resolution. The mode generation fidelity begins to decrease with the drive amplitude when the drive amplitude exceeds 1. The diffraction efficiency can be increased from (65\%) to a maximum of (75\%) increasing the drive amplitude from 1 to 1.3, although this comes at the cost of mode generation fidelity, which decreases from (70\%) to (66\%). Each point in the plot corresponds to 100 simulations, with the lines indicating the mean value, while the shaded regions correspond to $\pm 1$ standard deviation. In Fig. \ref{fig:response} (b), we simulated 100 sets of different possible RF waveforms for each of a set of resolutions ranging from 10 \textmu{}m to 1 mm. We see that at resolutions exceeding 100 \textmu{}m, we obtain high fidelity (>90\%) and efficiency (>80\%). The predicted value of the diffraction efficiency for large pixel sizes is in agreement with typical values (>80\%) specified for commercial AOMs for a single-frequency drive. For resolutions above 300 \textmu{}m, the fidelity can be as high as 99\% on average.

\begin{figure}[htbp]
\centerline{
\includegraphics[width=15cm]{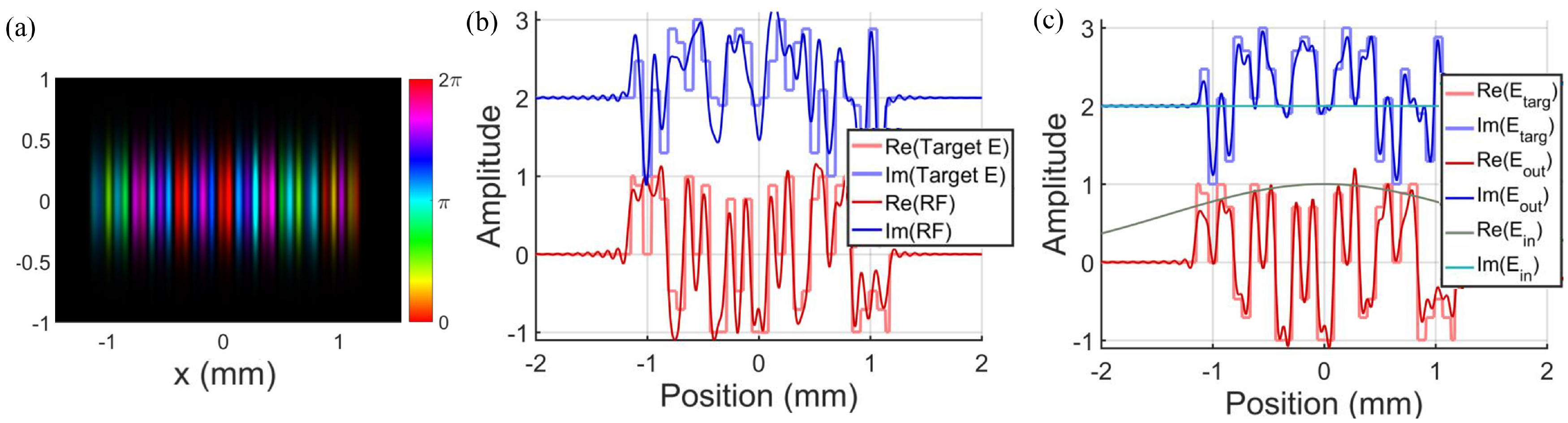}}
\caption{(a) Complex output field of a Zadoff-Chu sequence. (b) Comparison of the target spatial field and the optimized RF signal. (c) Input field and comparison of the target field to the output light field.}
\label{fig:sim_1Dseq}
\end{figure}

Fig. \ref{fig:sim_1Dseq} shows the generation of a Zadoff-Chu\cite{beyme2009efficient} sequence with 32 pixels, which approximates a mode with random phase and amplitude fluctuations. Orthogonal Zadoff-Chu sequences are used as the illumination patterns in single-pixel imaging, in which configurable SLMs are used to code the illumination with a set of spatial patterns, and the detection is performed with a single detector. Since the cyclically shifted versions of a Zadoff-Chu sequence are orthogonal to one another, this orthogonality property can further speed up frame rates in single-pixel imaging, because we don't need to wait for the acoustical wave to fully transit across the light beam before generating the next orthogonal illumination pattern, but rather just the transit time of the acoustical wave across one pixel of the Zadoff-Chu sequence. In our case, using the Zadoff-Chu sequence could speed up SPI by 32x. Fig. \ref{fig:sim_1Dseq}(a) shows the output field in which the colours and brightness correspond to the phase and the amplitude, respectively. Fig. \ref{fig:sim_1Dseq}(b) is a comparison between the target field and the optimized RF signal. The nonlinear effect, the nonideal response for sharp phase changes, makes the optimal RF signal and target light field different in their details, even though their overall structure is very similar. Fig. \ref{fig:sim_1Dseq}(c) compares the output light field after the AOM and the expected target field. The yellow and purple lines represent the real and imaginary parts of the input light field, where we can achieve 90\% fidelity with 75 \textmu{}m resolution. When comparing our approach with a rotating mask for SPI\cite{PhysRevA.78.061802}, we find that the rate at which new frames can be acquired equals the motion speed of the mask (or sound) relative to the spatial resolution of the illumination pattern. Thus, for the rotating disk method, assuming a pixel size of approximately 10 \textmu{}m and a speed of 1 m/s, it takes $\sim$10 \textmu{}s to get a new "frame". For our AO-SLM, the pixel size is $\sim$ 75 \textmu{}m, and the motional speed is $\sim$ 3.5 km/s; we obtain each new frame in about 21 ns, 50 times faster than a rotating disk approach for SPI.

\subsection{Single-pixel imaging}

\begin{figure}[htbp]
\centerline{\includegraphics[width=12cm]{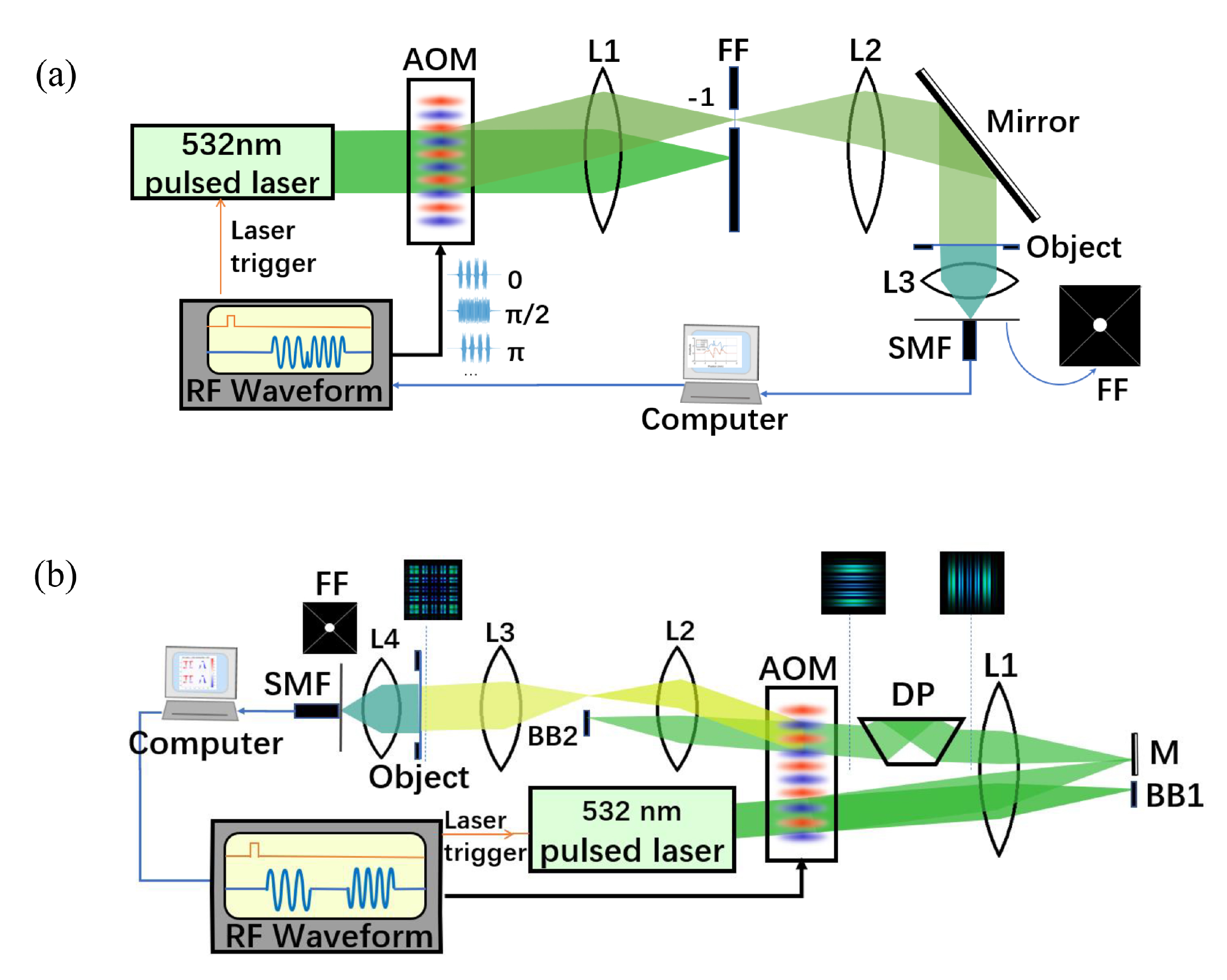}}
\caption{(a) Schematic of 1D single-pixel imaging. (b) Schematic of 2D single-pixel imaging. FF denotes the far field, SMF denotes the single-mode fiber, DP denotes the denotes the Dove prism, M denotes the mirror, BB denotes an opaque beam block, and L1, L2, L3 and L4 are optical lenses.}
\label{fig:spi}
\end{figure}

Fig. \ref{fig:spi} (a) shows the schematic for one-dimensional single-pixel imaging. Laser pulses are generated and sent through the AOM, where a portion is diffracted, replicating the spatial pattern produced by the RF waveform from an arbitrary waveform generator (AWG). We select the $'-1'$ order diffracted light in the far field (FF) within a 4-f imaging system and let it illuminate the complex field object, $S(\vec{r})$.

We then measure the single-point intensity at the center of the Fourier plane of the object plane, which can be expressed as

\begin{equation}
I_k =  \left| {P_k(\vec{r})\odot S(\vec{r})d\vec{r}}_{\vec{\kappa}=0} \right|^2 = \left| \int_\Omega{P_k(\vec{r})S(\vec{r})d\vec{r}} \right|^2
\label{eq:Ik}
\end{equation}
where  $\vec{\kappa}$ denotes the spatial frequency vector and $\vec{\kappa} = 0$ represents the center point of the Fourier plane, $\odot$ denotes the element-wise product at $\vec{\kappa} = 0$, $P_k(\vec{r})$ is one of the spatially modulated illumination patterns, and $\int_\Omega d\vec{r}$ denotes integration over the region of interest $\Omega$. In this way, the phases of both sample and illuminations are preserved.
In this work, we use the Zadoff-Chu sequence basis as the measurement matrix and three-step phase-shifting for phase imaging\cite{liu2018single}. By introducing a reference field and phase shifts in the complex structured illumination, $P_{k,\phi} = (e^{i\phi} \cdot Z_k + 1)/2$ where $Z_k$ being one of Zadoff-Chu sequences, the single-point intensity can be rewritten in a the phase-shifting interferometry form
\begin{equation}
I_{k,\phi} = \left| \int_\Omega{\frac{1}{2}(e^{i\phi} \cdot Z_k + 1)\cdot S(\vec{r})d\vec{r}} \right|^2 = \frac{1}{4}\left| e^{i\phi} \cdot \int_\Omega{ Z_k(\vec{r}))S(\vec{r})d\vec{r}} +\int_\Omega{ S(\vec{r})d\vec{r}}\right|^2 \\
=\frac{1}{4}\left| e^{i\phi} \cdot s_k +r \right|^2
\label{eq:I_k,phi}
\end{equation}
where $r= \int_\Omega{S(\vec{r})d\vec{r}}$ is a constant, which is the reference beam for the interferometry, $s_k = \int_\Omega{Z_k(\vec{r})S(\vec{r})d\vec{r}}$ is the integration of the complex field of the sample coded by the illumination pattern. With three-step phase-shifting, $\phi = 0, \pi/2, \pi$, the complex measurement field can be represented as
\begin{eqnarray}{}
y_k = (I_{k,0}-I_{k,\pi})+i(2I_{k,\pi/2}-I_{k,0}-I_{k,\pi}) = r\cdot s_k^*
\label{eq:yk}
\end{eqnarray}
where $s_k^*$ denotes the complex conjugate of $s_k$. 

We can simplify these expressions into vector form as follows: a sample with complex field $\boldsymbol{S}\in \mathbb{C}^N$ is illuminated by a spatially modulated pattern $\boldsymbol{P}\in \mathbb{C}^{M\times N}$, and $\boldsymbol{Y} = \boldsymbol{P}\cdot \boldsymbol{S}$ is the measurement field corresponding to $\boldsymbol{P}$, $\boldsymbol{Y}\in \mathbb{C}^M$. The sample field can be reconstructed via the corresponding inverse transform:
\begin{equation}
\bm{S} = \bm{P}^{-1}\bm{Y}
\label{eq:SPI}
 \end{equation}

\begin{figure}[htbp]
\centerline{\includegraphics[width=12cm]{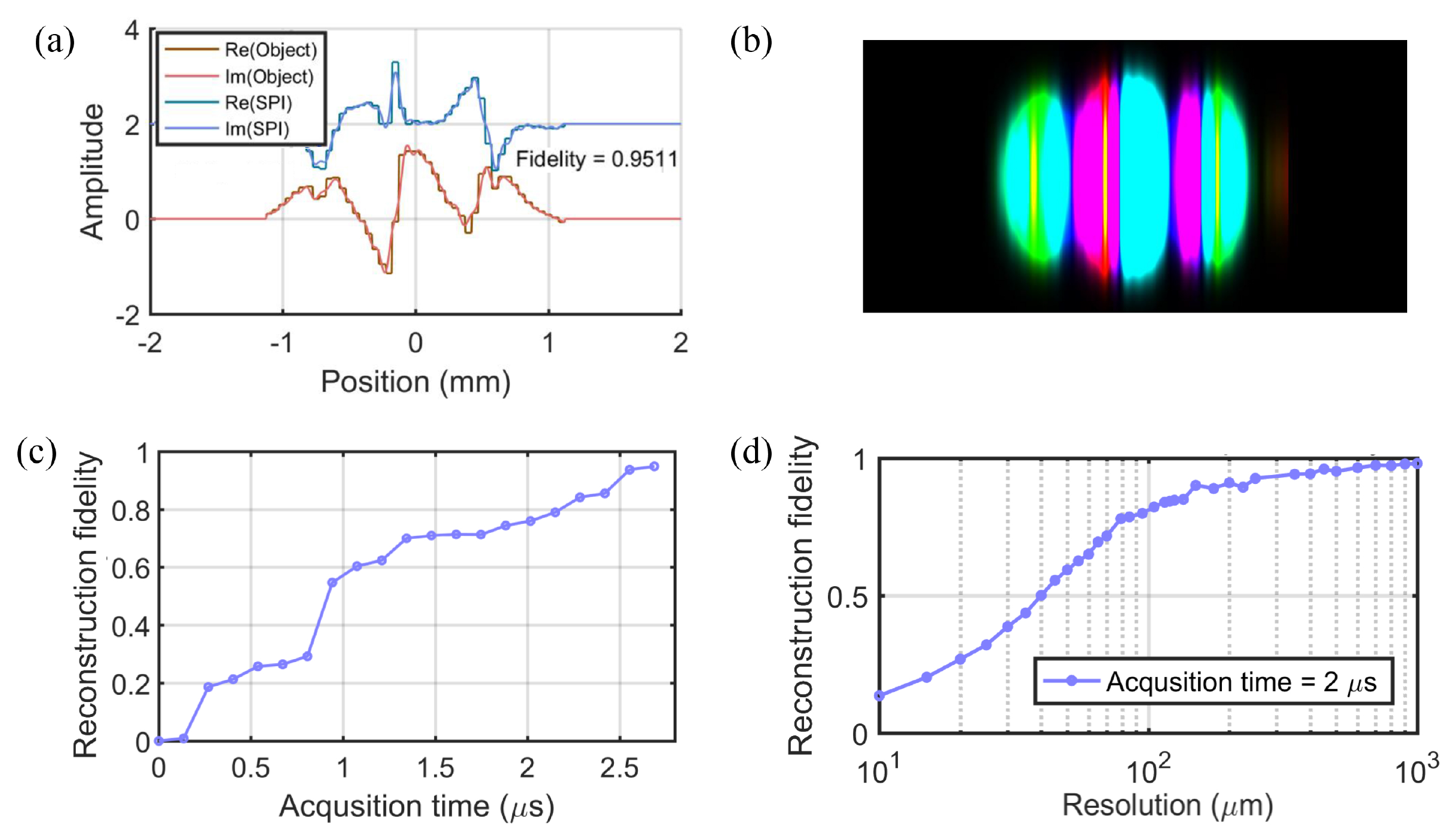}}
\caption{1D Single-pixel imaging results. (a) The comparison of the real and imaginary parts of Single-pixel imaging result and the object. (b) Complex field recovered by SPI, with hue corresponding to phase and brightness corresponding to intensity. (c) Fidelity versus acquisition time. (d) Fidelity of random object with different scale resolutions. The imaging is in all cases performed with the Zadoff-Chu sequence optimized at 75 \textmu{}m resolution.}
\label{fig:1D results}
\end{figure}

The one-dimensional single-pixel imaging simulation results using the optimized Zadoff-Chu sequence from Fig. \ref{fig:sim_1Dseq} are shown as Fig. \ref{fig:1D results} (a), (b). The fidelity is $95\%$ with 75 \textmu{}m resolution. Fig. \ref{fig:1D results} (c) demonstrates the fidelity versus acquisition time. The pattern refresh time, the time it takes for the RF wave to move 75 \textmu{}m forward, is about 21 ns for 3.6 km/s acoustic velocity. Therefore, recovering a 32 pixel complex signal requires only 2 \textmu{}s, which is three times the transit time because of the use of three-step phase-shifting Zadoff-Chu sequence. Fig. \ref{fig:1D results} (d) shows the fidelity of the arbitrary phase and amplitude object under different resolutions.

By designing a double-pass system as shown in Fig. \ref{fig:spi} (b), we can realize two-dimensional single-pixel imaging. Comparing to the one-dimensional system, the $-1$ order diffracted light is reflected back, and passes through a Dove prism, where the light field is rotated 90 degrees. The light then passes through the AOM a second time on a different location of AOM, where it is modulated in the second dimension. After an identical 4-f system, the structured light illuminates the 2-D complex object, and is detected in the same way as in the 1-D system.
\begin{figure}[htbp]
\centerline{\includegraphics[width=10cm]{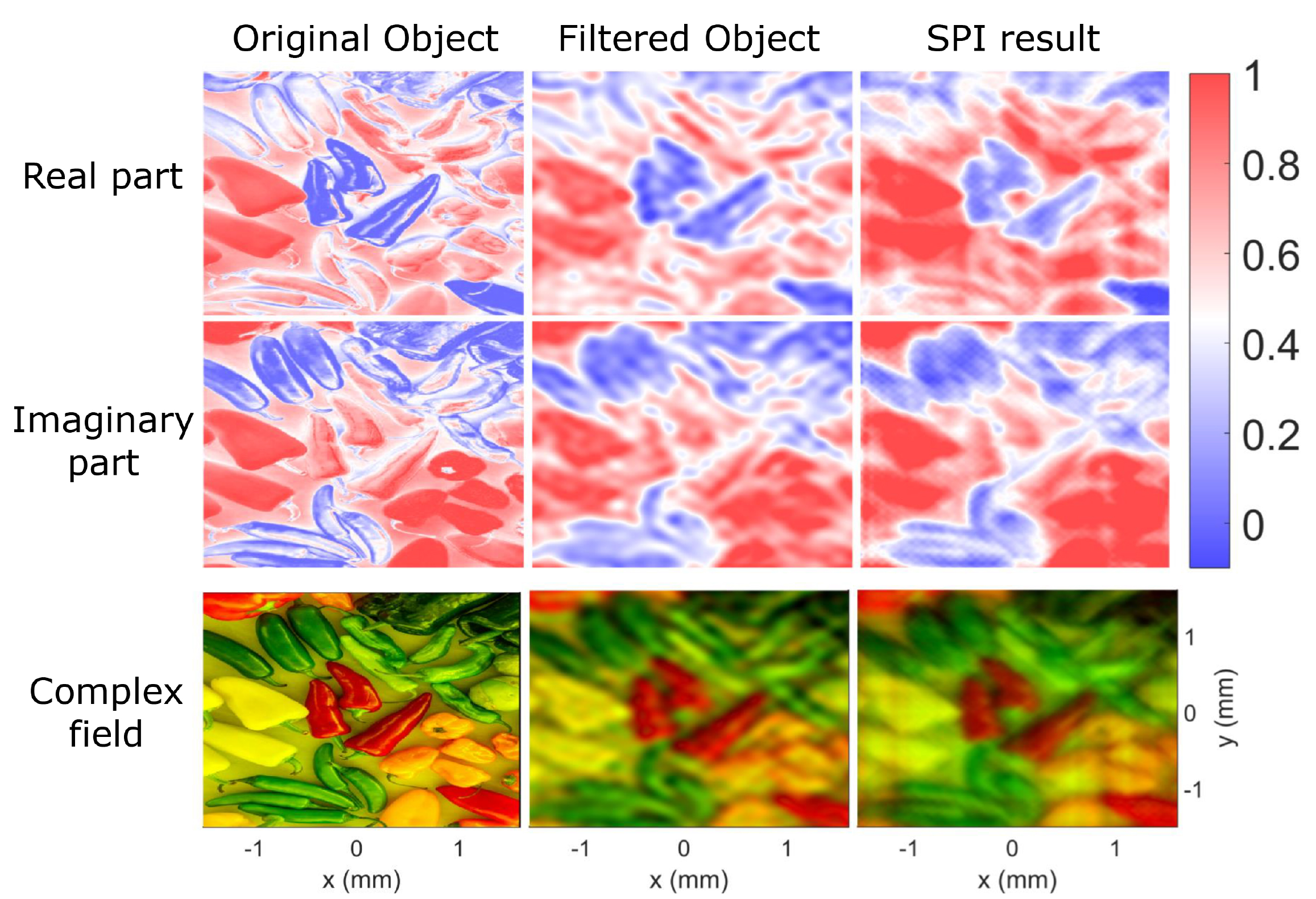}}
\caption{Two-dimensional single-pixel imaging results.}
\label{fig:2D results}
\end{figure}

We use the combination of two cascaded 1-D Zadoff Chu basis to generate a 2-D orthogonal basis. The measurement rate still depends on the time it takes for sound waves to travel one pixel, which is about 28 ns for the 100 \textmu{}m resolution we employ for the 2D simulations. For a set of 32 $\times$ 32 Zadoff-Chu pattern, it takes about 86 \textmu{}s to perform a fully-sampled complex field reconstruction, which is faster than the refresh rate for a single frame DMD and traditional SLM. Since the illumination pattern is scanned continuously by the motion of the sound through the crystal, to improve the reconstruction quality and reduce the impact of the sampling grid, we perform two samplings of each pixel, corresponding to a 14 ns sampling interval. The results of 2D single-pixel imaging are shown in Fig. \ref{fig:2D results}. The object is a 600$\times$600 pixel "peppers" image, and we use the real/imaginary components of the field to represent the red/green channels of the image. The simulation step size is dozens of times smaller than the pixel size of a Zadoff-Chu pattern, so we can use a high-resolution object to test our imaging result. 

The first column is the original object, while the middle column shows the spatially filtered object, which incorporates the loss of high-frequency information due to the spatial filter in the far-field plane that selects the first-order diffraction. The upper resolution limit of single-pixel imaging is always determined by the cutoff frequency of the spatial filter, regardless of the specific technique used to generate the illumination pattern. Therefore it is reasonable to compare the SPI result with the image after the spatial filter. The third column is the recovery result by SPI, and the fidelity of this reconstruction relative to the filtered image is about 98$\%$. 

\section{Experimental Results}
\begin{figure}[htbp]
\centerline{
\includegraphics[width=12cm]{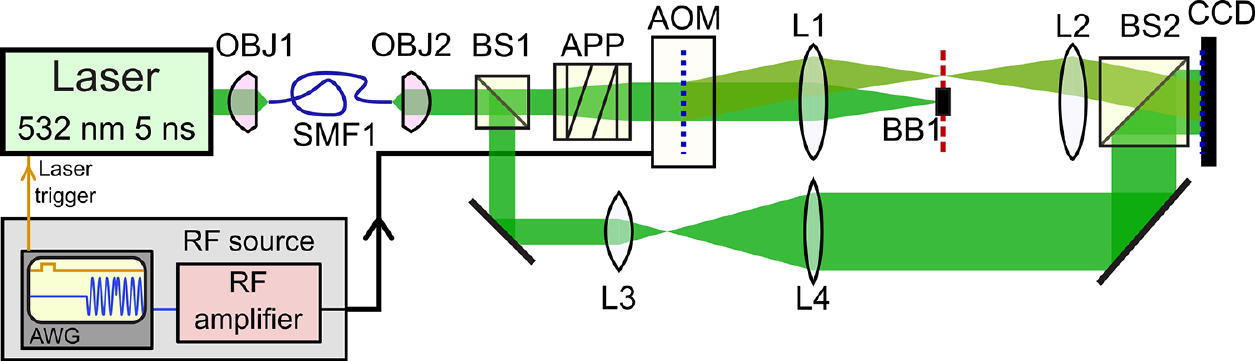}}
\caption{Experimental setup for employing an AOM as a SLM. }
\label{fig:setup}
\end{figure}

To verify the feasibility of our scheme, we built an experimental setup. The schematic of the experimental apparatus is shown in Fig. \ref{fig:setup}. Light pulses are generated and are coupled into a single-mode fiber to produce a consistent spatial mode. The light is then coupled back into free space and split into two beam paths: signal and reference. Signal light passes through the AOM, where a portion is diffracted, replicating the spatial pattern produced by the RF waveform from an arbitrary waveform generator (AWG). By analyzing the interference between the two beams captured by the CCD camera, we recover both the amplitude and phase information of the generated spatial mode, a technique referred to as off-axis holography. 

\begin{figure}[htbp]
\centerline{
\includegraphics[width=13cm]{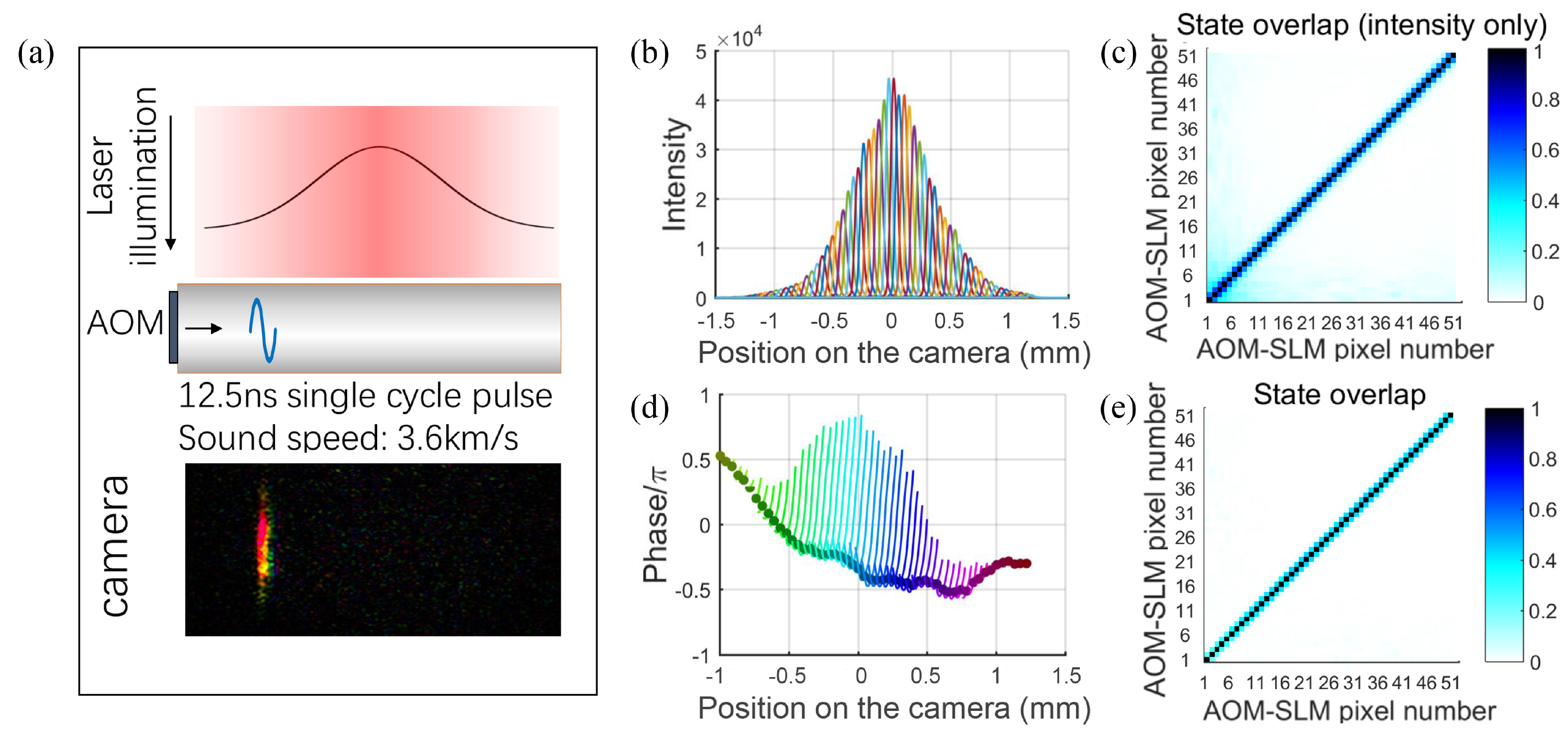}}
\caption{Calibrating system response with scanning single RF cycle. (a) The model of the calibration process. (b) The intensity response of individual AO-SLM pixels. (c) The state overlap between output fields of different AO-SLM pixels including intensity information only. (d) The phase response of individual AO-SLM pixels. (e) The state overlap between output fields of different AO-SLM pixels including phase infromation.}
\label{fig:Calibration}
\end{figure}

To calibrate the system response of our AO-SLM, we use a 12.5 ns single cycle RF pulse scanning the whole beam as shown in Fig. \ref{fig:Calibration} (a). Each RF cycle correspond to a 45 \textmu{}m slit (12.5 ns RF pulse at the propagation speed of 3.6 km/s) on the image plane which can be regarded as a single programmable pixel. The intensity and phase response of these pixels is shown in Fig. \ref{fig:Calibration} (c) and (d). The envelope of the intensity distribution of the input beam is Gaussian, which is consistent with the input light being a Gaussian beam. The phase distribution is not very constant. These steep lines are caused by the delayed response of the AOM. We use the phase values at solid dots position in Fig. \ref{fig:Calibration} (c) to compensate the RF signal, that is, to set the initial phase offset for the corresponding pixel. The analysis plots in Fig. \ref{fig:Calibration} (c) and (e) show the cross-talk between the different RF channels (pixels on the reconstructed light plane). In Fig. \ref{fig:Calibration} (c) where only the light field intensity is considered in computing the pixel cross-talk, adjacent pixels have significant overlap of approximately 70\%. In contrast, in Fig. \ref{fig:Calibration} (e), the state overlap including phase information shows the pixels are in fact nearly orthogonal when phase information is considered.

\begin{figure}[htbp]
\centerline{
\includegraphics[width=15cm]{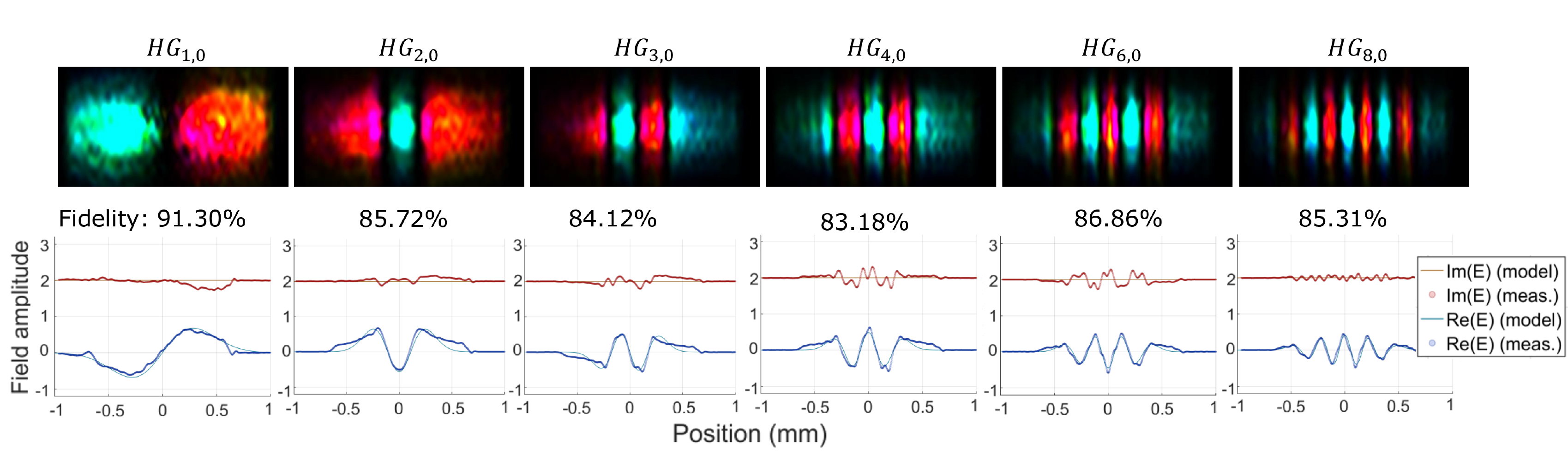}}
\caption{Experimental results for generation of HG modes.}
\label{fig:HGmodesExp}
\end{figure}

\begin{figure}[htbp]
\centerline{
\includegraphics[width=14cm]{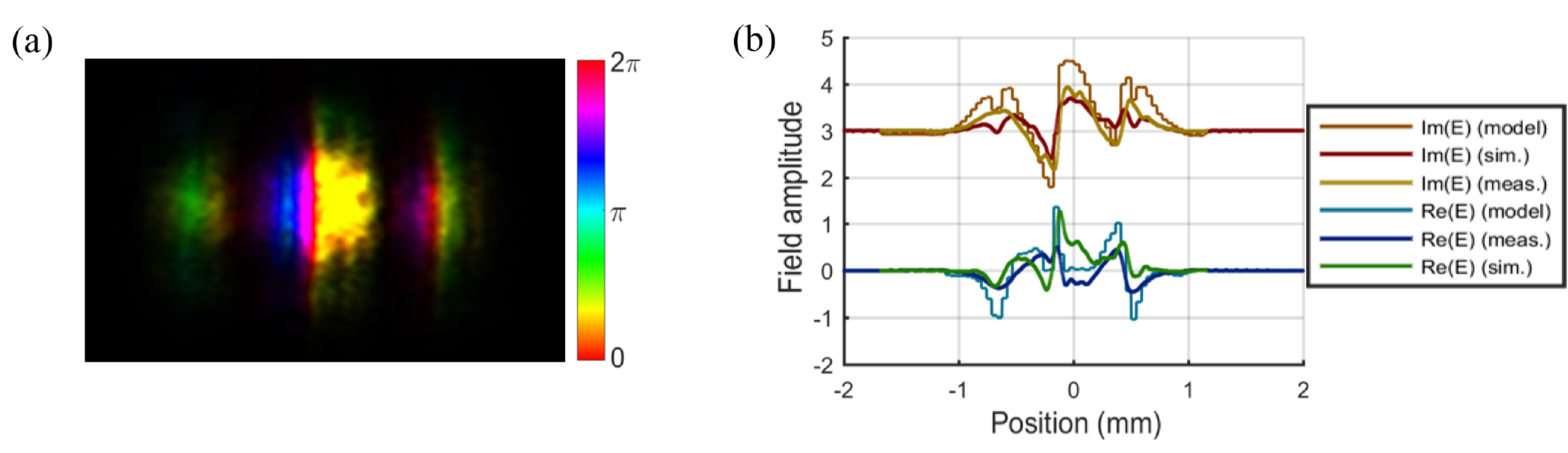}}
\caption{The experimental result for generation of an arbitrary mode. (a) The complex field of an arbitrary mode; (b) Comparison of the imaginary and real parts between simulation and experimental result.}
\label{fig: Exp arbitrary mode}
\end{figure}

Fig. \ref{fig:HGmodesExp} shows the experimental results for generating approximations to HG modes at 45 \textmu{}m resolution. The hue represents phase, and the brightness represents intensity. We generate the HG$_{1,0}$ mode with fidelity 91\%, and fidelities of around 85\% for modes HG$_{n,0}$ for n={2,3,4,6,8}. The average overlap between the target and realized mode is 85\%. We also generated a specific mode as shown in Fig. \ref{fig: Exp arbitrary mode}. The imaginary and real parts of simulation and experimental result and the target model is shown in Fig. \ref{fig: Exp arbitrary mode} (b). The fidelity of experimental result is 86.3\%, and the overlap between the simulation and experimental result is 93.4\%. These data are collected without optimizing the RF waveform. Although the fidelity of the experimental results is somewhat lower than that of the simulation, we can still say that our simulation model can describe the nonlinear effect of the AOM, but some additional systematic differences between the experimental setup and the simulation configuration result in disagreement between simulation and experiment. In simulation, we used the RF waveform optimization to improve the agreement between the generated and target spatial modes, but we found the RF waveforms optimized through computer simulations not improve experimental performance significantly. Additional measurements will be necessary to fully understand and overcome the nonlinear effects in the experimental system. It is likely that RF waveform optimization could be implemented experimentally by using the measured optical field as an input to the optimization algorithm.


\section{Conclusion} 
The effective pixel size we realize with our AO-SLM is approximately 50 \textmu{}m, currently limited by the response time of the piezo-electric-actuator driving the crystal. Across the 2 mm width of the probe beam, we can have up to 40 pixels with independently controllable amplitudes and phases. The fidelity of mode generation can be as high as 85\% for 50 \textmu{}m spatial resolution after RF waveform optimization. The fast refresh rate of the modulation pattern (approaching 20 MHz) speeds up single-pixel imaging, which can be performed at a rate of up to 11.6 kHz for the reconstruction of a 32$\times$32 pixel complex object, thousands of times faster than using crystal SLMs or DMD\cite{Liu2018Fast,Welsh2013Fast,Phillips2016Adaptive}. The AO-SLM inherits the broadband performance of AOMs, allowing patterning light over a wavelength range of 100s of nm.

Our results suggest that using an AOM as an SLM is a promising approach for generating complex spatial modes of light for pulsed and short-wavelength lasers, which could damage traditional DMDs and SLMs. By designing a double-pass system, where the light beam is rotated by 90 degrees before the second pass through the crystal, we can achieve two-dimensional spatial light modulation. The refresh rate is only limited by the transit time of the sound waves across the beam waist, with sub-microsecond switching possible. Since the AO-SLM is a diffractive device, the efficiency of mode generation is reduced by the amplitude modulation of the RF signal since some undiffracted light is blocked. This can be solved by passing the beam through the AOM a second time. In that case, high efficiency would be attained by phase-only modulation, with arbitrary mode generation arising from the cascading of two-phase masks\cite{Scholes:20}.

\section*{Funding}

Canada Research Chairs (501100001804); Office of Naval Research (100000006): 2204-202-2023940, N00014-17-1-2443, N00014-19-1-2247;
Natural Sciences and Engineering Research Council of Canada (501100000038).

\section*{Acknowledgments}

R.B. acknowledges support through the Natural Sciences and Engineering Research Council of Canada, the Canada Research Chairs program, US Office of Naval Research Awards 2204-202-2023940, N00014-17-1-2443, and N00014-19-1-2247. 
B. B. acknowledges the support of the Banting postdoctoral fellowship.
X. L. acknowledges the financial support from China Scholarship Council and Shanghai Science and Technology Development Funds.

\section*{Disclosures}

The authors declare that there are no conflicts of interest related to this article.

\bibliography{aoslm}

\end{document}